# Magnetic properties of optimized cobalt nanospheres grown by Focused Electron Beam Induced Deposition (FEBID) on cantilever tips

Soraya Sangiao[1,2,*], César Magén[1,2,3], Darius Mofakhami[4], Grégoire de Loubens[4] and José María De Teresa[1,2,5]

*[1]Laboratorio de Microscopías Avanzadas (LMA), Instituto de Nanociencia de Aragón (INA), Universidad de Zaragoza, 50018 Zaragoza, Spain.*

*[2]Departamento de Física de la Materia Condensada, Universidad de Zaragoza, 50009 Zaragoza, Spain.*

*[3]Fundación ARAID, 50018 Zaragoza, Spain*

*[4]Service de Physique de l'Etat Condensé, CEA, CNRS, Université Paris-Saclay, 91191 Gif-sur-Yvette, France.*

*[5]Instituto de Ciencia de Materiales de Aragón (ICMA), CSIC - Universidad de Zaragoza, 50009 Zaragoza, Spain.*

[*]Email: Soraya Sangiao – sangiao@unizar.es

**Abstract**

In this work, we present a detailed investigation of the magnetic properties of cobalt nanospheres grown on cantilever tips by Focused Electron Beam Induced Deposition (FEBID). The cantilevers are extremely soft and the cobalt nanospheres are optimized for Magnetic Resonance Force Microscopy (MRFM) experiments, which implies that the cobalt nanospheres must be as small as possible while bearing high saturation magnetization. It is found that the cobalt content and the corresponding saturation magnetization of the nanospheres decrease for nanosphere diameters below 300 nm. Electron holography measurements show the formation of a magnetic vortex state in remanence, which nicely agrees with magnetic hysteresis loops performed by local magnetometry showing negligible remanent magnetization. As investigated by local magnetometry, optimal behavior for high-resolution MRFM has been found for cobalt nanospheres with diameter of ≈200 nm, which present atomic cobalt content of ≈83 at% and saturation magnetization of $10^6$ A/m, around 70% of the bulk value. These results represent the first comprehensive investigation of the magnetic properties of cobalt nanospheres grown by FEBID for application in MRFM.

## Introduction

Through the local decomposition of magnetic precursor molecules by the action of an incoming electron beam, a wide range of functional magnetic nanostructures has been grown in last years by the Focused Electron Beam Induced Deposition (FEBID) technique [1,2]. The extensive list of nanostructures includes: (a) planar deposits in the shape of Hall bars for sensing purposes [3–6]; (b) magnetic nanopillars for functionalization of tip cantilevers with applications in Magnetic Force Microscopy (MFM) [7–10] and Magnetic Resonance Force microscopy (MRFM) [11]; (c) planar nanowires for application in magnetic domain-wall conduits [12,13], in logic circuits [14,15], in dense memory arrays [16] and for superconducting-vortex-lattice pinning [17]; (d) three-dimensional nanowires for magnetic domain-wall studies [18,19] and for remote magnetomechanical actuation [20], dots for magnetic storage [21] and catalytic purposes [22], polygonal shapes for micromagnetic studies [23,24] and spin-ice investigations [25], nanoconstrictions and nanocontacts for domain-wall pinning [26] and Andreev-reflection studies [27], etc. The growth of such numerous types of 2D and 3D magnetic nanostructures has been possible thanks to the main virtues of the FEBID technique: arbitrary design of the beam-scan path [28], high resolution provided by the fine electron beam spot [29], tuning of growth parameters (beam dwell time, precursor flux, etc.) [30,31] and flexibility in the type of substrate used (rigid or flexible, flat or curved, conductive or insulating) [32].

An important aspect to consider in the growth by FEBID is the metal content, which is generally linked to the functionality of the deposit. In the case of magnetic deposits grown by FEBID, the metal content can be finely tuned in various ways. The beam current [7,33], the beam dwell time [30], the precursor flux [5], the beam voltage [34] and the substrate temperature [35,36] have been found to be relevant parameters to tune the metal content in magnetic deposits. However, some constraints exist, which impede to grow arbitrary shapes with arbitrary metal content. In general, the difficulties increase when the target is to grow very small structures (smaller than 100 nm) with high metal content. Another strategy to increase the metal content and/or change the microstructure consists in the application of post-growth purification steps [37–39]. In order to avoid the surface oxidation of the magnetic nanostructures, the use of protective shells have been found to be very effective [31,40].

In the present work, we challenge the growth of cobalt nanospheres by FEBID for application in MRFM. MRFM is a quantitative magnetic characterization technique that exploits the tiny magnetic forces appearing between a magnetic tip and a magnetic sample for the investigation of spin dynamics at the nanoscale [41]. This near field scanning probe technique allows magnetic resonance imaging (MRI) with nanometer spatial resolution and extreme spin sensitivity [42] and the investigation of spin-waves at the sub-micron scale [43–45]. In these applications, very strong field gradients from the magnetic probe [46] and ultra-soft cantilevers [47] are required. Therefore the magnetic probe should be precisely located at the apex of the cantilever, be as small as possible to gain spatial resolution, and have as high magnetization as possible to maximize the MRFM signal [48]. Moreover, a spherical shape is beneficial to minimize hysteresis effects and makes any quantitative analysis more easy [49]. These requirements imply the optimization of the FEBID growth in order to obtain cobalt spheres sufficiently small but at the same time showing high metal content in order to present high saturation magnetization.

# Results and Discussion

## Sample growth and characterization

In FEBID, the precursor gas molecules are delivered onto the substrate surface by means of a nearby gas-injection system and the focused electron beam is scanned on the surface. The precursor gas molecules are dissociated by electron beam irradiation, creating a deposit with the same shape of the beam scanning. The cobalt nanospheres were grown by FEBID under 5 kV electron beam voltage. For cobalt nanospheres with diameter above 150 nm, an electron beam current of 1.6 nA was used; whereas for growing smaller nanospheres an electron beam current of 0.4 nA was chosen. The main growth parameters used for the cobalt nanospheres reported here are listed in Table 1.

In order to synthesize nanospheres, we have taken advantage of the point-like nature of the growth surface, i.e. the apex of the cantilever (Olympus BioLever, around 30 nm in size). We have scanned the beam over a circular area centered on the apex of the cantilevers and varied the radius of the circular area being scanned and the time of beam scanning to obtain the different targeted diameters. The diameter of the circular area is constant during the growth of each nanosphere and equal to approximately 75% of the targeted diameter. Beam shift together with live imaging were used to ensure that the circular area being scanned is always centered on the apex of the cantilever. Along several optimization experiments, we have chosen the optimal radius and time of the circular area being scanned for growing nanospheres with desired diameters. As shown in figure 1 for three different cobalt nanospheres grown by FEBID, we are able to fabricate cobalt nanospheres with desired diameter by optimizing the radius and the time of circular area being scanned on the apex of the cantilever. The nanospheres required growth times ranging from 2 to 6 s. In the present work we have fabricated cobalt nanospheres with diameter ranging from 500 nm down to 90 nm, with good spherical geometry and smooth surface.

| Φ Nanospheres diameter | $V_{beam}$ (kV) | $I_{beam}$ (nA) | % at. Co for Φ = 400 nm |
|---|---|---|---|
| Φ < 150 nm | 5.0 | 0.4 | 93 (2) |
| Φ ≥ 150 nm | 5.0 | 1.6 | 91 (2) |

**Table 1:** Growth parameters used for the cobalt nanospheres in the present study. In the last column, the cobalt content (atomic percent) of a cobalt nanosphere of 400 nm in diameter grown under the reported conditions is given.

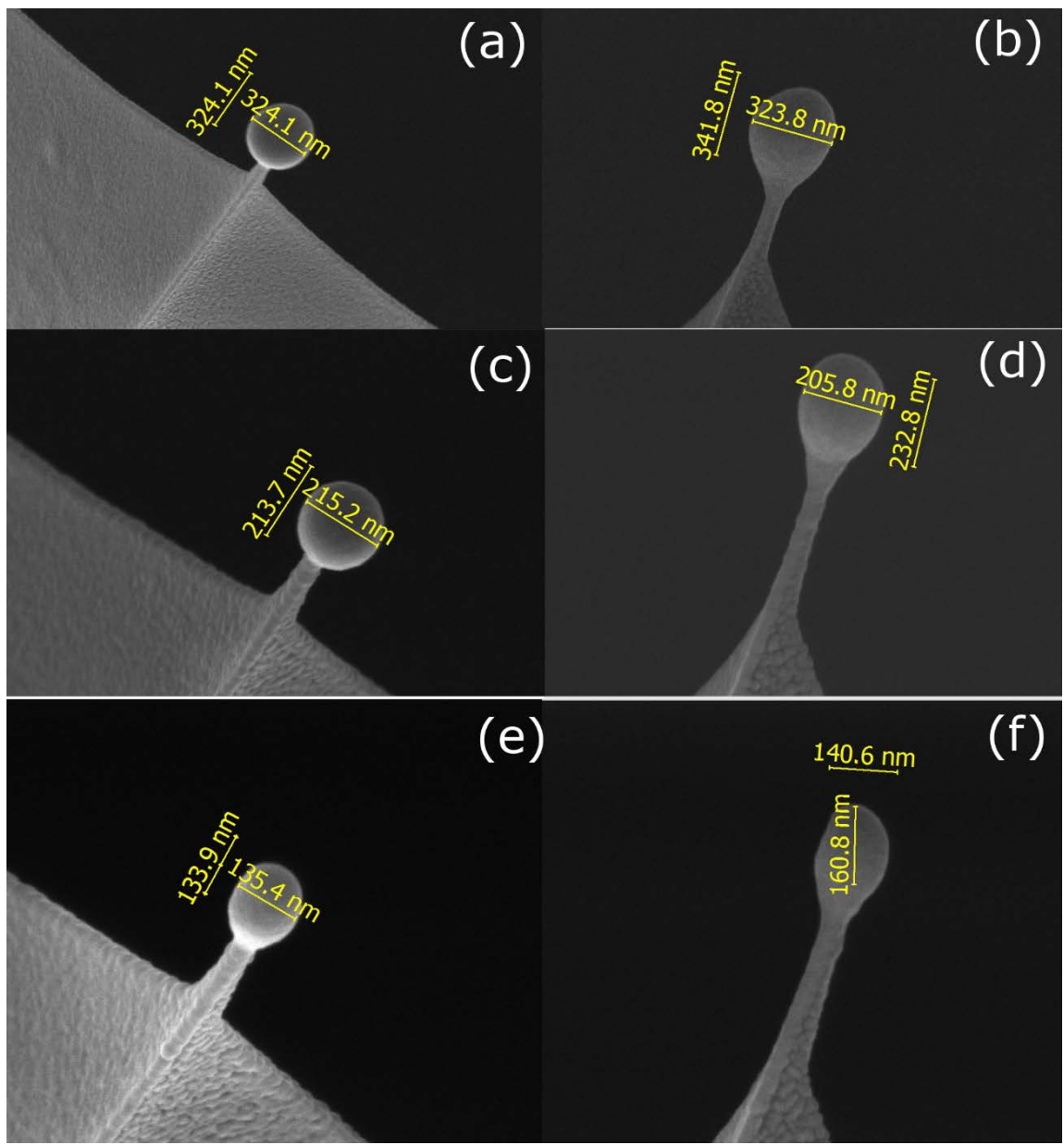

**Figure 1:** SEM micrographs showing the dimensions of the grown cobalt nanospheres. Top view and front view of cobalt nanospheres of 325 nm (a, b), 215 nm (c, d) and 135 nm (e, f) in diameter.

We have studied the cobalt content of the nanospheres grown by FEBID at the apex of cantilevers by Energy Dispersive X-ray Spectroscopy (EDS) to check the evolution of the metal content with the nanospheres' diameter, as it changes the surface to volume ratio of the nanosphere. First, we have grown nanospheres of 400 nm of diameter using the two different sets of growth parameters reported on separate rows of Table 1, which are the most appropriate for growing cobalt nanospheres with diameters either lower or higher than 150 nm, respectively. The obtained cobalt contents for both nanospheres so grown, listed in the last column of Table 1, are very similar to each other and the difference between both values, 93% at. Co and 91% at. Co respectively, is below the experimental accuracy. As shown in Figure 2, the cobalt content, in atomic percent, decreases as the diameter of the nanosphere decreases, down to the minimum value found of 60 % at. Co for the smallest nanosphere of 90 nm in diameter. The optimized cobalt content, ~92% at., is obtained only for nanospheres with diameter higher than 400 nm. We attribute the decrease in the cobalt content for diameters below 400 nm to the natural surface oxidation of the cobalt nanospheres, which occurs in a spherical shell with an outer radius equal to the radius of the particular nanosphere and a thickness of approximately 5 nm. Another explanation, besides native surface oxidation, for the decrease in cobalt content for diameters below 400 nm could be a change in the growth mode, as previously reported in 3D cobalt nanowires grown by FEBID [31]. In 3D cobalt nanowires the growth occurs in a radial mode for wire diameter higher than 120 nm, with resulting higher Co content than wires with diameter smaller than 80 nm, which grow in the linear mode.

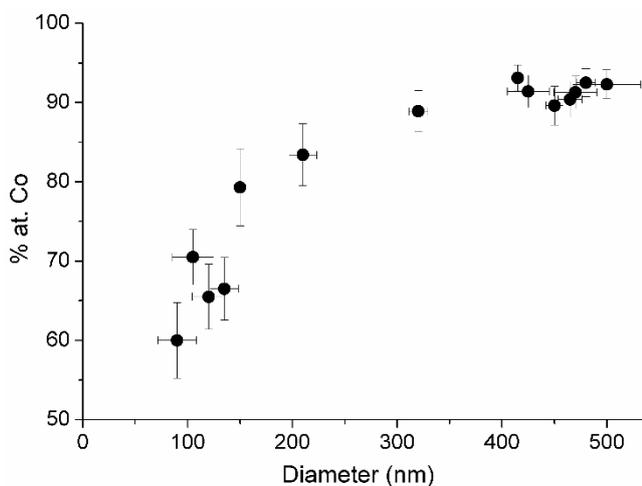

**Figure 2:** Composition of the cobalt nanospheres, as measured by EDX, as a function of their diameter.

In order to analyze the chemical composition of the cobalt nanospheres by Electron Energy Loss Spectroscopy (EELS) in scanning transmission electron microscopy mode (STEM) and their local magnetic properties by Electron Holography in a transmission electron microscope (TEM), the specimens were prepared for TEM observation in a specific geometry. Firstly, the cantilever pyramid tip is cut by focused ion beam (FIB) milling and lifted-out by a micromanipulator. Then, the cantilever tip is welded onto a TEM copper grid by a FIB-induced Pt deposition, as illustrated in Figure 3(a). Then, the FEBID cobalt nanosphere is grown at the apex of the cantilever, following the same procedure as described above. Figures 3(b) and (c) display the SEM micrographs of the two cobalt nanospheres studied by STEM-EELS and Electron Holography, once grown at the apex of cantilevers already attached to the TEM grid.

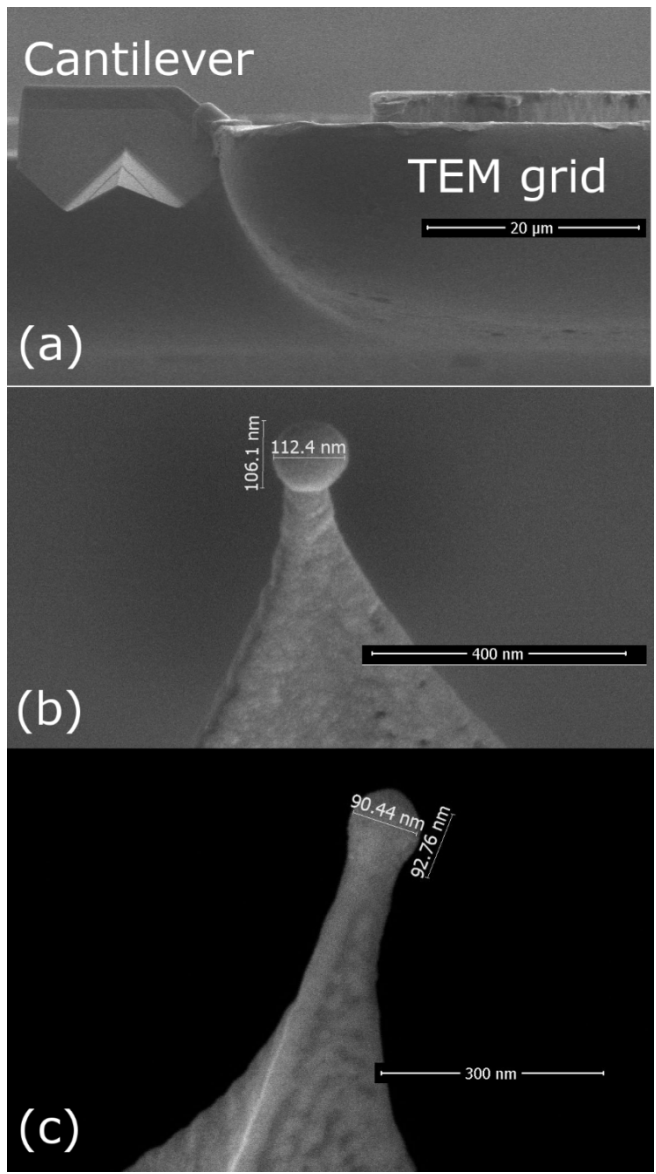

**Figure 3:** SEM micrographs of the cobalt nanospheres grown on cantilever tips for STEM-EELS and Electron Holography experiments. (a) Cantilever pyramid tip welded to a TEM grid. (b, c) Cobalt nanospheres grown by FEBID on the apex of the cantilever already attached to the TEM grid. The diameter of the nanospheres shown is 110 nm (b) and 90 nm (c).

The morphological and compositional properties of the cobalt nanospheres grown by FEBID have been confirmed by local chemical mapping of selected nanospheres of diameters 110 nm (see Figure 4b) and 90 nm (see Supporting Information) performed by STEM-EELS. These quantitative maps reveal, first of all, that the deposits are not perfect spheres attached to the tip. On the other hand, they appear to be partially stuck into the tip of the cantilever, in particular the smallest sphere. For this morphology, secondary electrons that cause the decomposition of the precursor are emitted all around the tip; thus, at the early stages of the growth, cobalt atoms wrap around the tip of the pyramid. A coloured chemical map, including the relative compositions of Co (red), O (green) and C (blue), the only chemical elements detected in the nanospheres, is shown in figure 4b for the nanosphere of 110 nm in diameter. This chemical map can be analyzed quantitatively, as displayed in figure 4c, obtaining a net Co content at the center of the nanosphere of about 80% at. with respect to the total composition of Co, C and O. A remarkable oxidation layer is observed, extending approximately 6 nm. This agrees nicely with previous reports on Co-FEBID, which have confirmed this layer to be non-ferromagnetic [50]. Furthermore, a thin layer containing carbon and oxygen of about 7 nm is formed due to contamination before and during the electron beam irradiation in the TEM experiment. As a result, the average diameter of cobalt under the oxidation layer and possible contaminant extends to 100 nm out of the 110 nm of the whole sphere diameter.

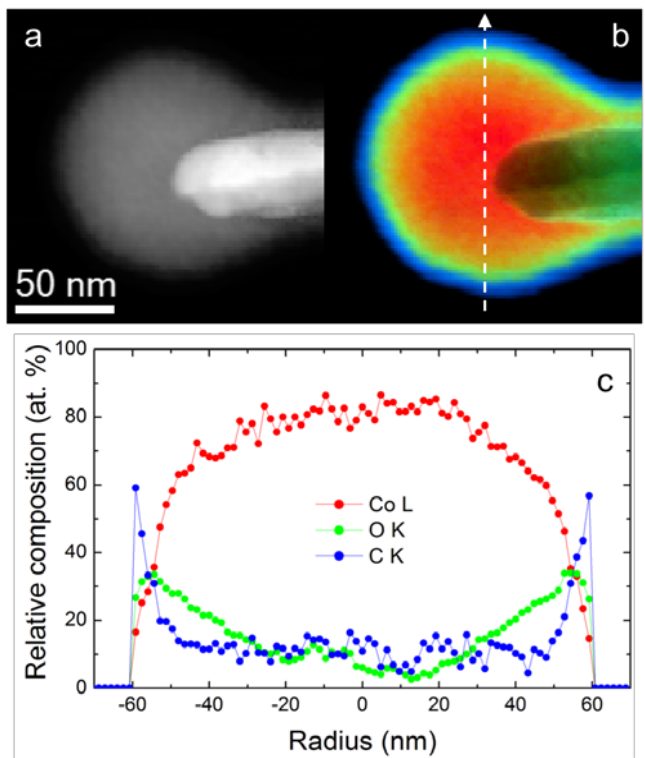

**Figure 4:** STEM-EELS compositional analysis of the cobalt nanosphere with 110 nm of diameter. (a) Reference image in Z contrast. (b) Colored chemical map, including the relative compositions of Co (red), O (green) and C (blue). (c) Compositional line profile extracted along the white arrow in (b).

## Local Magnetic Characterization by Electron Holography

The remanent magnetic state of the two Co nanospheres with approximately 110 nm and 90 nm in diameter (see figures 3b and 3c) has been imaged by off axis Electron Holography in a TEM [51].

Figure 5 illustrates the hologram acquisition and retrieval of the magnetic induction flux distribution in the nanosphere with 110 nm grown on the cantilever tips, (see the analysis of the nanosphere with 90 nm diameter in the Supporting Information). Figure 5(a) shows the bright field image of the Co nanosphere overlapped with the interference fringe pattern of the hologram, revealing a significant amount of contamination which has not disappeared after standard $Ar/O_2$ plasma cleaning procedures. Furthermore, the holograms before and after (not shown) reversing the object show how contamination builds up during the experiment, what affects the quantitativenes of the technique (particularly for the smaller sphere of 90 nm, shown in the Supporting Information).

The electrostatic and magnetic contributions to the phase shift retrieved from the holograms analysis are shown in Figure 5(b) and Figure 5(c), respectively. In particular, the magnetic contribution illustrated in Figure 5(c) is shown in terms of the cosine of 12 times the magnetic phase, giving rise to a fringe pattern that corresponds to the distribution of magnetic induction flux lines produced by the magnetic object. As a result, the nanosphere presents a nearly circular closure domain of magnetic induction, circulating counterclockwise, with the in-plane magnetization decreasing while approaching the center of the nanosphere. No in-plane stray fields are observed. This magnetic induction geometry corresponds to a counterclockwise vortex state of undetermined polarity (it is compatible with the magnetic flux leaking at the center of the sphere both into or out of the image plane). Quantitative values of the in-plane magnetic induction can be extracted by estimating the local thickness of the sample using the electrostatic phase image. This is done by assuming that the area of maximum phase around the center of the sphere corresponds to a nominal thickness of 110 nm (the contribution of the contamination is ignored, assuming that the contribution of the carbon contamination layer to the average mean inner potential of the object is reduced). Using this "thickness" image, the absolute in-plane magnetic induction map can be determined. A line profile of the net in-plane magnetic induction distribution along the white arrow in Figure 5(c) is displayed in Figure 5(d). This magnetic induction profile matches again that of a vortex state in which the maximum magnetic induction values are observed at the outer regions of there; these values diminish while approaching the center of the sphere due to the rotation of the magnetization out-of-plane until the in-plane magnetic induction is nearly zero around the center of the nanosphere, which corresponds to the vortex core. The maximum value of the magnetic induction is approximately $1.1 \pm 0.1$ T, which agrees nicely with previous magnetization values determined for similar nanodeposits, such as vertical nanowires [31]. In both cases, deposition condition gives rise to magnetization values reduced with respect to bulk values due to the moderate purity of the deposit and the diminution of the effective magnetic volume due to the formation of an non-magnetic oxide surface layer [50].

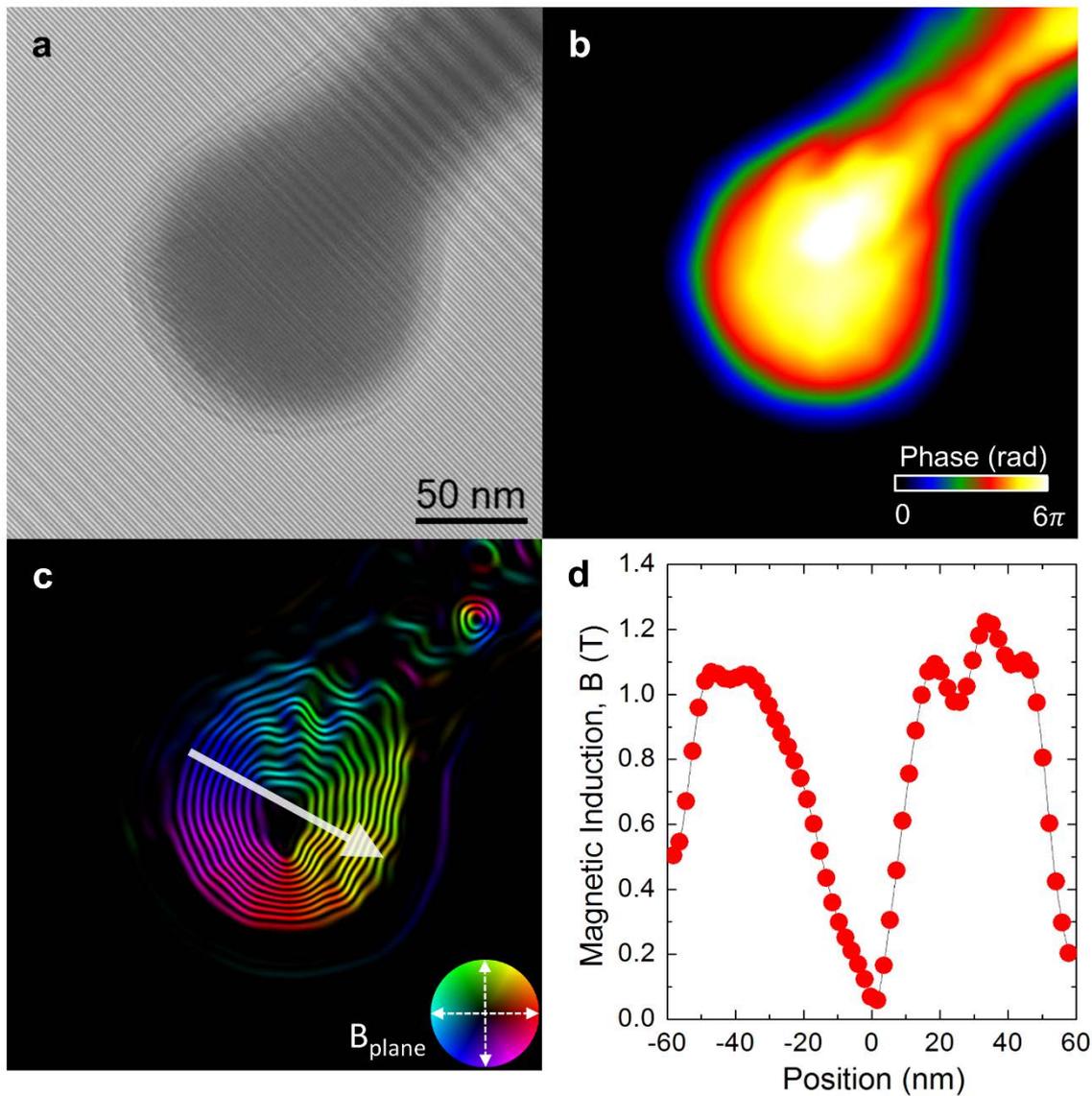

**Figure 5:** Electron holography of one cobalt nanosphere with a diameter of 110 nm. a) Electron hologram of the object. b) Electrostatic phase image, $\varphi_E$. c) Colored representation of the in-plane magnetic induction flux lines, represented as $cos(12\varphi_M)$. The inset represents the color scale of the magnetic induction orientation in arbitrary units, where the position of a color relative to the center of the circle corresponds to the orientation of the magnetic induction. d) Profile of the in-plane component of the magnetic induction vector as measured along the white arrow c), where the position reference is taken at the minimum of the magnetic induction.

# Magnetization measurements of the cobalt nanospheres

In order to measure the magnetization of the cobalt nanospheres, we take advantage of them being attached at the end of very sensitive force sensors to perform cantilever magnetometry. The mechanical resonance frequency of the cantilever is monitored as a function of the applied magnetic field while it is positioned in a strong field gradient created by a cylindrical magnetic microwire [11], (see inset of Figure 6(a) and experiemental section for details on the set-up). Due to the low stiffness (spring constant k = 6 mN/m) and high quality factor (2000 < Q < 4000 under vacuum) of the cantilever, its frequency accurately probes the magnetic force produced by the field gradient on the nanosphere. In the experimental conditions, the cantilever frequency shift is directly proportional to the magnetization of the cobalt nanosphere (see equation 1 in experimental section), which allows simple extraction of its hysteresis curve. This is shown in Figures 6(a) and (b) for a cobalt nanosphere having a diameter of 500 nm. The nanosphere is fully saturated above 0.6 T, and its magnetization decreases quite linearly with the field below this value to become negligible in the remanent state. These magnetometry data also allow us to quantitatively extract the magnetization of the nanosphere [11]. From the maximal relative variation of the cantilever frequency, 1.2% in Figure 6 (a), and knowing the cantilever spring constant and the second spatial derivative of the magnetic field ≈ (1.5 ± 0.3) $10^9$ T/m$^2$ in which the measurements are operated, one can estimate the magnetic moment of the 500 nm diameter cobalt nanosphere to be (1 ± 0.2) $10^{-13}$ A.m$^2$. Divided by the volume of the nanosphere, this yields a saturation magnetization $M_s$ of 1450 ± 300 kA/m, which compares well to the bulk value of cobalt at room temperature (1400 kA/m). To check the consistency of this estimate and obtain a better accuracy, one can also use the value of the saturation field of the nanosphere. For a perfect sphere without crystalline anisotropy, it is only governed by demagnetizing effects and equal to $\mu_0 M_s/3$. This saturation field is accurately determined from a series of measurement similar to the one presented in Figure 6 (a) by varying the distance from the source of the field gradient. The saturation field for the 500 nm diameter cobalt nanosphere is found to be 0.58 ± 0.01 T, which yields $M_s$ = 1385 ± 25 kA/m assuming a perfect spherical shape, in very good agreement with the previous estimate.

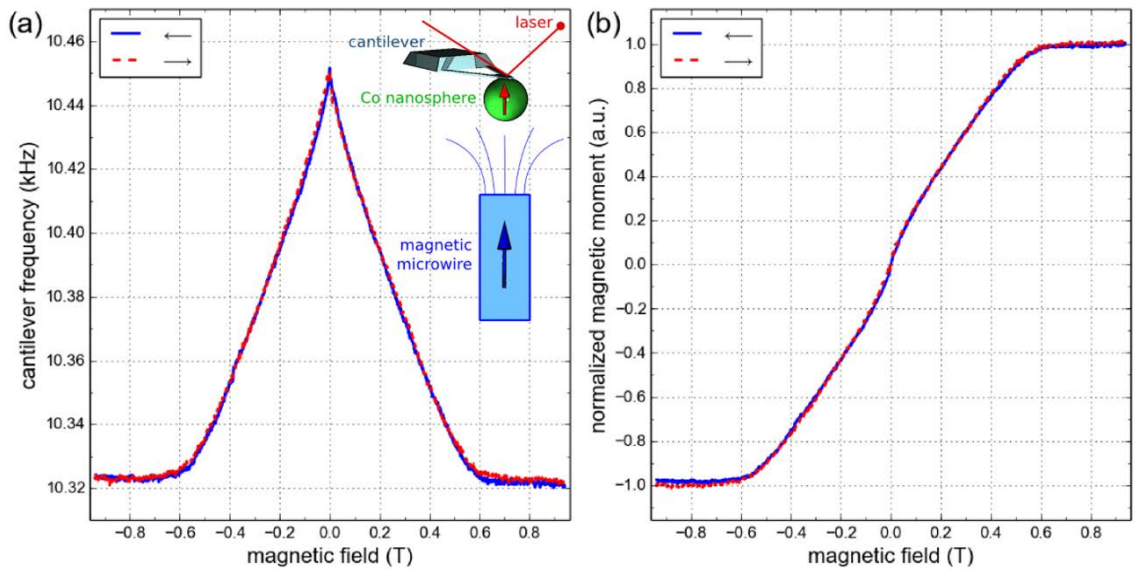

**Figure 6:** Cantilever magnetometry of a 500 nm diameter cobalt nanosphere. (a) Raw data of the cantilever frequency vs. applied magnetic field. Inset: Sketch of the cantilever magnetometry set-up. (b) Extracted magnetization curve.

We have repeated these magnetometry measurements for different magnetic nanospheres of varying diameters. The experimental results are reported in Figure 7, where the dependence of the saturation magnetization upon the diameter of the cobalt nanosphere is displayed. It is found that for diameters above 300 nm, the saturation magnetization of the nanosphere is close to bulk cobalt, in good correspondence with the behavior of the cobalt content, which remains close to 90 at% in that range of diameter. Below 300 nm, the saturation magnetization of the nanosphere quickly drops, similarly to the decrease of the cobalt content observed in Figure 2. By extrapolating, one would find that the saturation magnetization vanishes for a cobalt content below 50 at%. Interestingly, 200 nm diameter nanospheres still have a magnetization of about 1000 kA/m, which for MRFM application represents the best compromise between spatial resolution and sensitivity.

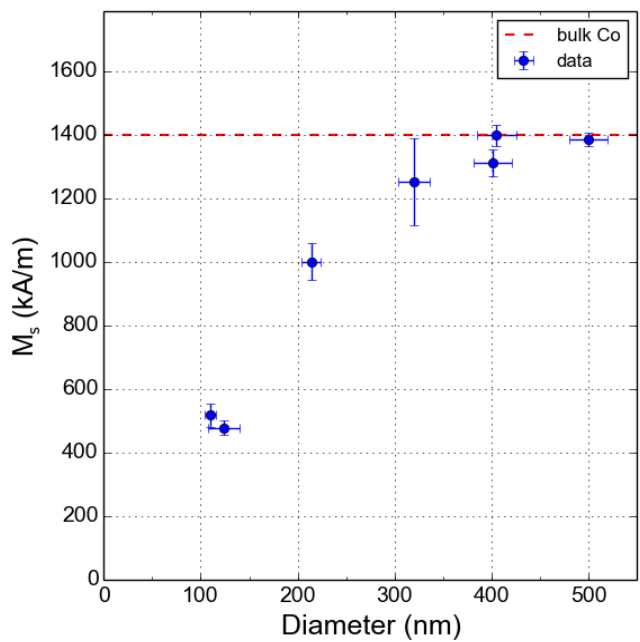

**Figure 7:** Saturation magnetization of the cobalt nanospheres as a function of their diameter.

# Conclusion

Summarizing, we have presented here a comprehensive characterization of the chemical and magnetic properties of cobalt nanospheres grown on the apex of MRFM cantilever by FEBID. EDS analysis of the atomic Co content shows a maximum of 94 at% for nanospheres with diameters higher than 400 nm and a decrease in the Co content for smaller nanospheres.

Quantitative chemical composition analysis by STEM-EELS on a cobalt nanosphere of 110 nm in diameter has shown a relative Co content of 80 at% and has revealed the presence of a native oxidation spherical shell of 6 nm in thickness. Precise characterization of the remanent magnetic state has been performed by electron holography on the cobalt nanospheres of 110 nm in diameter. The in-plane magnetic induction geometry corresponds to a counterclockwise vortex state.

As investigated by local magnetometry, optimal behavior for high resolution MRFM has been found for cobalt nanospheres with diameter around 200 nm, which present atomic cobalt content of 83 at% and saturation magnetization of about $10^6$ A/m, 70% of the cobalt bulk value. This study constitutes the first detailed characterization of the magnetic properties of cobalt nanospheres grown by FEBID for application in MRFM experiments.

## Experimental

Samples have been grown by FEBID using the following parameters: $V_{beam}$ =5 *kV*, $I_{beam}$ = 0.4 nA for diameter below 150 nm and 1.6 nA for diameters above 150 nm, beam spot diameter = 8.8 *nm (0.4 nA) / 17.6 nm (1.6 nA)*, precursor temperature = 27ºC, chamber base pressure ≈ 1.2 x $10^{-6}$ mbar, chamber growth pressure ≈ 3.5 x $10^{-6}$ mbar. EDS experiments have been performed using a beam voltage of 5 kV.

Cantilever magnetometry measurements were performed at room temperature using the set-up described in refs. [11,49]. The source of the field gradient is a millimeter long, 16 µm diameter cylinder of CoFeNiSiB alloy, with a saturation magnetization of 510 kA/m. The cobalt nanosphere is positioned at a distance between 5 µm and 20 µm from the top surface of the cylinder to perform the measurements. A standard laser deflection technique is used to monitor the displacement of the cantilever. Its resonance frequency is tracked using a piezoelectric bimorph and a feedback electronic circuit based on a phase lock loop. The relative frequency shift due to the force acting on the magnetic moment *m* of the cobalt particle is:

$$\frac{\Delta f_c}{f_c} = -\frac{m}{2k}\left(\frac{\partial^2 B_z}{\partial z^2}\right)_{|z_0} \quad (1)$$

where *k* is the cantilever spring constant, $B_z$ the vertical component of the magnetic field from the cylinder, and $z_0$ the equilibrium position of the particle in the field gradient.

STEM-EELS chemical mapping and quantification was carried out at an acceleration voltage of 300 kV in a probe-corrected FEI Titan 60-300 equipped with a high brightness field emission gun (X-FEG), a CEOS corrector for the condenser system and a Gatan Tridiem 866 ERS image filter/spectrometer. The EELS acquisition was performed with a convergence angle of 25 mrad, a collection semi-angle around 60 mrad, an estimated beam current of 160 pA and an exposure time of 30 ms/pixel. The chemical composition was determined by the standard method of integrated intensity elemental ratios implemented in Gatan's Digital Micrograph software package, using the carbon K, oxygen K and cobalt $L_{2,3}$ edges. No further correction for thickness effects was applied".

Off-axis Electron Holography has been carried out in an FEI Titan Cube 60-300 equipped with a Schottky field emission gun (S-FEG), a CEOS corrector for the objective lens and a motorized electrostatic biprism. The experiments have been performed in aberration-corrected Lorentz mode, with the objective lens switched off and the corrector aligned to compensate the spherical aberration of the Lorentz lens and achieve a spatial resolution of around 1 nm. Electron holograms of ~20% contrast have been obtained with a biprism excitation of 160 V, an overlap region of about 500 nm and an acquisition time of 8 s. The electrostatic phase shift ($\varphi_E$) and the magnetic phase shift ($\varphi_M$) are retrieved by recording two holograms for each object, the second one with the object flipped with respect to the original orientation. In this way, the magnetic contribution contained in the holograms changes sign, while the electrostatic contribution due to the mean inner potential remains unchanged. Once the phases are extracted from both holograms, their subtraction produces a pure magnetic phase shift image and the magnetic induction can be calculated as:

$$\boldsymbol{B}_{x(y)} = \frac{\hbar}{e \cdot t} \frac{\partial \varphi_M}{\partial y(x)} \qquad (2)$$

Where ℏ is the reduced Planck's constant, e is the electron charge and t is the thickness of the sample.

Visualization of the magnetic state of the Co nanospheres is performed by calculating the cosinus of a multiple of the magnetic phase shift, $cos(n\varphi_M)$, which produces sets of fringes parallel to the magnetic induction flux. Absolute values of magnetic induction are calculated by estimating the local thickness of the object from normalization of $\varphi_E$ to the total diameter of the nanosphere.

## Supporting Information

Chemical compositional maps by STEM-EELS and electron holography experiments have been performed on the sphere with a diameter of 90 nm following the exact same procedure described in the main text, and these results are described and illustrated in the supporting information.

## Acknowledgements

This work was supported by Spanish Ministry of Economy and Competitivity through projects No. MAT2014-51982-C2-1-R, MAT2014-51982-C2-2-R and MAT2015-69725-REDT (including FEDER funds), by the Aragon Regional Government, and by EU (Cost Action CM1301- CELINA and ESTEEM2 projects). Experimental help in the sample growth by Laura Casado (LMA-INA) is acknowledged.